\documentclass[10pt,twocolumn,preprintnumbers,amsmath,amssymb,nofootinbib
,superscriptaddress]{revtex4}
\usepackage{graphicx,graphics}

\newcommand{\be}{\begin{equation}}  
\newcommand{\ee}{\end{equation}}  
\newcommand{\bea}{\begin{eqnarray}}  
\newcommand{\eea}{\end{eqnarray}}  
\newcommand{\ba}{\begin{array}}  
\newcommand{\ea}{\end{array}}

\def\ie{\hbox{\it i.e.}{}}	  
\def\eg{\hbox{\it e.g.}{}}

\def\VEV#1{\left\langle #1\right\rangle}

\def\beq{\begin{equation}}  
\def\eeq{\end{equation}}  
\def\bea{\begin{eqnarray}}  
\def\eea{\end{eqnarray}}  
\def\half{\frac{1}{2}}  
  
\def\bq{\begin{quote}}  
\def\eq{\end{quote}}

\def\half{\frac{1}{2}}       
\relax  
\newskip\humongous \humongous=0pt plus 1000pt minus 1000pt

\newif\ifdtup

\def\oldreffmt#1{\rlap{[#1]} \hbox to 2\parindent{}}

\def\figfmt#1{\rlap{Figure {#1}} \hbox to 1in{}}  
  
\def\ie{\hbox{\it i.e.}{}}	  
\def\eg{\hbox{\it e.g.}{}}

\def\beq{\begin{equation}}  
\def\eeq{\end{equation}}  
\def\bea{\begin{eqnarray}}  
\def\eea{\end{eqnarray}}  
\def\half{\frac{1}{2}}  
  
\def\bq{\begin{quote}}  
\def\eq{\end{quote}}

\def\half{\frac{1}{2}}       
\relax  

\newdimen\tdim  
\tdim=\unitlength  
\def\bar{\overline}

\begin{document}

\preprint{FERMILAB-PUB-18-076-T}

\title {Inertial Symmetry Breaking
}

\author{Christopher T. Hill}
\email{hill@fnal.gov}
\affiliation{Fermi National Accelerator Laboratory\\
P.O. Box 500, Batavia, Illinois 60510, USA\\$ $}

\date{\today}

\begin{abstract}

We review and expand upon recent work 
demonstrating that  Weyl invariant theories can 
be broken ``inertially,''  which does not
depend upon a potential. This can be understood in a general
way by the ``current  algebra'' of these theories, independently of
specific Lagrangians.
Maintaining the exact Weyl invariance in a renormalized quantum theory
can be accomplished by renormalization conditions that
refer back to the VEV's of fields in the action. 
We illustrate the computation
of a Weyl invariant Coleman-Weinberg potential that breaks a $U(1)$
symmetry together with scale invariance.
\footnote{Invited talks delivered at: ``The Great Solar Eclipse Conference,''
Aug. 20, 2017, Columbia, MO, USA; ``Collider Physics and the Cosmos,''
Galileo Galilei Institute, Oct. 9, 2017, Arcetri, Florence, Italy; 
Physics Colloquium, Departamento de Fisica Teoricam
Universidad de Valencia, Oct. 18, 2017, Burjassot, Valencia, Spain; Fermilab Seminar, Winter 2017. } 
\end{abstract}

%\pacs{14.80.Bn,14.80.-j,14.80.-j,14.80.Da}
\maketitle

\bigskip

\section{Introduction}

\bigskip

The discovery of the Higgs boson,
a fundamental, pointlike, scalar field, unaccompanied by a natural 
custodial symmetry, has led many authors to turn to scale symmetry
in search of
new organizing principles.
Weyl symmetry \cite{Weyl} provides a natural setting for scalars with gravity
and may provide a foundational symmetry
for physics  \cite{shapo}\cite{GB}\cite{authors}\cite{Englert:1976ep}\cite{S3}\cite{FHR}.
Though Weyl symmetry is technically
distinct from scale (diffeomorphism) symmetry (as we
discuss in Section II), the two have many features in
common, and we will refer casually 
to ``Weyl'' and ``scale'' symmetries 
interchangeably. 

Weyl symmetry, like many candidate symmetries seen in nature, must be broken.
Breaking is generally treated spontaneously, implemented via  potentials. 
However, for Weyl  symmetry
such potentials cannot contain any explicit mass scales. Thus,
one cannot sculpt a conventional ``Mexican hat'' potential since
 a  $-m^2\phi^2$
term is disallowed. However, potentials that can break
Weyl symmetry have general properties that
are both interesting and restrictive. 

For example, if we have a multi-field 
scale invariant potential, $W(\phi_i)$, then scale invariance implies
$\sum \phi_i\delta W/\delta\phi_i =4W$.
It follows that,
if the fields develop nontrivial vacuum expectation values (VEV's),  $\VEV{\phi_i}$, 
such that  $\delta W/\delta \VEV{\phi_i}=0$, then necessarily $W(\VEV{\phi_i})=0$, \ie,
the cosmological constant is zero at the symmetry breaking minimum.  
It immediately follows that there must
be a flat direction, since if 
$\VEV{\phi_i}\rightarrow \lambda \VEV{\phi_i}$, then  $W(\lambda\VEV{\phi_i})
=\lambda^4W(\VEV{\phi_i})=0$, hence the potential energy is zero along
the flat direction.
A dilaton then arises as the Nambu-Goldstone boson of spontaneously broken
Weyl symmetry. Such potentials will be considered in Sections IV and V.

Our main goal in the present note, however, is to 
summarize a new way to break Weyl and other
symmetries that {\em does not employ a potential}.
This
mechanism is implicit in many of the approaches taken to spontaneously generating
the Planck scale, but it has not to our knowledge been been codified prior to ref.\cite{FHR,Ferreira:2018itt}.
The resulting ``current algebra'' is interesting and powerful, allowing general
statements to be made for a wide class of 
Weyl invariant theories.
This is a direct consequence of the 
structure of the Weyl current and the presence of  
inflation in the  early universe. It unifies the inflationary
universe with the formation of the Einstein-Hilbert effective action
and the Planck scale itself. It can break other symmetries in
the gauge sector of the standard model and its extensions, and can lead to large
hierarchies. We call this mechanism {\em ``inertial spontaneous symmetry breaking''}
\cite{Ferreira:2018itt}.

Most authors generally
construct scale invariant theories, starting in the  ``Jordan frame,'' which
is manifestly scale invariant.
Typically one then performs
a Weyl transformation to the ``Einstein frame,'' where the conventional
mass scales (\eg, Planck mass, cosmological constant, etc.) are introduced
as parameters of the transformation. 
The ensuing inflationary dynamics is usually discussed in this Einstein frame,
where the dynamics is most familiar.

But,  various questions then arise: Where did the Einstein frame mass parameters, such
as $M_{Planck}$, come from?
If Weyl symmetry is spontaneously broken what is the ``order parameter''
of the broken phase? For example, in the standard model there is
a clear distinction between the symmetric (Higgs VEV zero) and broken (Higgs VEV nonzero)
phases.  What is the analogue distinction between the symmetric initial (Jordan frame) theory
and the ultimate Einstein frame theory? 

Presently we follow ref.\cite{Ferreira:2018itt,FHR} and observe that, by remaining in
the original Jordan frame and treating the full evolutionary dynamics there, one can observe
how the theory evolves and spontaneously breaks scale symmetry.  The key ingredient
is the Weyl scale symmetry  current, $K_\mu$.  

We find that the expansion of the 
universe in a pre-inflationary
phase drives  the current charge density, $K_0$, to zero. This is 
much as any conserved current charge density, such as the 
density of electric charge,
dilutes to zero by general expansion.  
However, the Weyl current, $K_\mu$, 
is the derivative of a scalar,  $K_\mu =\partial_\mu K$, 
and in particular $K_0 =\partial_0 K$.
We refer to $K$ as the ``kernel'' of the current.  
Hence, as the $K_\mu$ current density is diluted away, 
$K_0\rightarrow 0$,  the kernel $K$ therefore evolves to an arbitrary
constant:  $K\rightarrow \bar{K}$ constant.  
In a Weyl invariant theory this implies that scale symmetry is broken,
and a Planck mass
is generated that is $\propto \sqrt{\bar{K}}$.  

Hence,
any random initial values of the fields and their derivatives will 
inflate to a universe
that is then described with a Planck mass, and other mass scales.
$K$ plays the role of the symmetry breaking order
parameter and directly defines $M_{Planck}^2$. 
A potential may then be needed to engineer the
final vacuum, and determine the ratios of individual
fields $\VEV{\phi_i}$, yet it plays no direct role 
in the inertial Weyl symmetry breaking phenomenon.

With a little
thought,
one can guess the structure of the order parameter $K$.
Consider a set of $N$ scalar fields $\{\phi_i\}$.
The fields are non-minimally coupled
to gravity as $(-1/12) \sum_{i} \alpha_i\phi_i^2 R(g)$.
Therefore, if any of the $\phi_i$ should develop a VEV,
we would expect scale breaking, and a nonzero $K$. Hence,
we expect that the order parameter takes the form, $K\sim c\sum_i \phi_i^2$.
However, if any  $\phi_i$ has $\alpha_i = 1$, then we can
remove it from the action by a local Weyl transformation, absorbing it into the metric.
We therefore expect $K= c'\sum_i (1-\alpha_i) \phi_i^2$.
Indeed, we will find that $K_\mu = \partial_\mu K$,  with $c'=1/2$, by a Noether variation
of the Jordan frame theory under a Weyl transformation, thus confirming our guess.
More elegantly, we can find $K_\mu = \partial_\mu K$ and $D_\mu K^\mu = 0$ 
directly from the trace of the
Einstein equations in the Jordan frame, combined with the Klein-Gordon (KG)
equations.  We emphasize that $K_\mu$ is conserved ``on-shell,'' a consequence of 
both the Einstein and KG equations. 

The result that $K\rightarrow\bar{K}$ constant as the universe expands
implies that $N-1$ $\{\phi'_i\}$  will ultimately
satisfy the constraint  $\bar{K}=(1/2)\sum_i (1-\alpha_i) \phi'_i{}^2$.
The constrained $\{\phi'_i\}$ lie on an ellipsoid in field space.
This leaves one field unconstrained
that becomes the dilaton, and it is intimately
related to the $K_\mu$ current. If we perform a field redefinition,
\beq
\phi_i = \exp(\sigma/f)\phi_i'\qquad
 g_{\mu\nu}= \exp(-2\sigma/f) g'_{\mu\nu}.  
\eeq
we will find for the Weyl invariant action:
\bea
S(\phi, g) & =& S(\phi', g') \nonumber \\
& & \!\!\!\!\!  \!\!\!\!\!  \!\!\!\!\!   \!\!\!\!\! \!\!\!\!\!
+ \int \sqrt{-g'} \left( \partial_\mu K(\phi')\partial^\mu (\sigma/f) 
+ K(\phi') (\partial \sigma/f)^2 \right)
\eea
Now using the constraint that $K(\phi')=\bar{K}$ constant,
we have:
\beq
S(\phi, g) = S(\phi', g') 
+ \frac{1}{2}\int \sqrt{-g'} (\partial \sigma)^2 
\eeq
Here we identify $f^2=2\bar{K}$ so the dilaton is
canonically normalized.
From this we see that the dilaton, $\sigma$, describes a dilation
of the ellipse, and moves in field space orthogonally to the $N-1$  $\{\phi'_i\}$ fields.
 Hence the dilaton decouples
from everything except gravity (this holds true for fermions and gauge bosons
as well).

We further see that the current written in the unconstrained
fields is equivalent to one written in the constrained fields
by: $K_\mu = \partial_\mu K(\phi)= \partial_\mu (K(\phi') e^{2\sigma/f})$.
Hence in the broken phase (Einstein frame) limit $K(\phi')\rightarrow \bar{K}$ constant, 
$K_\mu \rightarrow 2\bar{K}\partial_\mu \sigma /f=f \partial_\mu \sigma$.
where $f = \sqrt{2\bar{K}}$.
This is as  we expect for a Nambu-Goldstone boson, \eg, the axial current of the pion takes 
the analogous form $f_\pi \partial_\mu \pi$.

Why is this formulation important?
Results following from the ``current algebra''
of Weyl invariant theories are general statements that are true,
independent of the specific structure of the Lagrangian. The particular structure
of $K_\mu$ and $K$ is independent of the form
of any scale invariant potential, but the detailed structure of $K$ does
depend upon the choice of the non-minimal
couplings (and also any higher derivative gravitational terms can modify the
simple forms we just discussed). The behavior of the current algebra will remain intact,
since $K_\mu=\partial_\mu K$ is conserved, 
but \eg, the ellipsoid defined by $\bar{K}$ could become a more general locus. 

The survival of the general feature of inertial breaking with
a stable goundstate, \eg, a stable $M_{Planck}$, requires that the
quantum theory does not break Weyl symmetry through loops.  
Here it
is important that one does not conflate the procedure of {\em regularization}, which generally
introduces arbitrary mass scales,
with {\em renormalization}, which introduces counterterms to define the final theory
and its symmetries.  Though it is convenient, one need not deploy a regulator that
is consistent with the symmetries of a renormalized theory. In fact,
such a regulator may not exist, though the
symmetries can
exist in the final renormalized theory.\footnote{ Even dimensional regularization cannot by itself
determine the chiral anomalies of a theory such as the standard model, 
which are ultimate positioned in
the relevant currents by the choice of counterterms.}
Furthermore, physics
should not depend upon the choice of regulator, and even
scalar masses in a theory with a momentum space cutoff can be viewed
as technically natural, as emphasized by Bardeen \cite{bardeen}.
The nonexistence of a symmetry in the regulator does not imply
the nonexistence of the symmetry in the renormalized theory: only anomalies
do. 

Presently we must confront the meaning of scale anomlies which could
seemingly appear
in quantum loops and spoil $D_\mu K^\mu = 0$.   
Scale anomalies occur in an effective theory and reflect the 
need for the specification of input scales,
usually at higher energies.  For example, the QCD scale anomaly reflects the need
for a definition of $g_3$ at some high energy scale, which can serve 
as the boundary condition of the running coupling $g_3(\mu)$. 
We can view the origin of $\Lambda_{QCD}$ in the low energy world
as ``predicted''
via the solution to the RG (Renormalization Group) equation, given a high
energy input parameter.
At one-loop order:
\beq
\label{QCD}
\frac{\Lambda_{QCD}}{M} = \exp\left(\frac{1}{8\pi b_0\alpha_3(M)}\right)
\eeq
where $b_0=-(11/3 - 2n_f/3)$. Here $\alpha_3(M)$ is the input value
of $g_3^2/4\pi$, the QCD coupling
defined at the scale $M$, and defines the
particular RG trajectory of the running $\alpha_3(\mu)$.
The low energy scale $\Lambda_{QCD}$ is ``inherited'' from
the implicit high energy scale $M$ by the RG evolution.
Here we have explicitly broken scale symmetry
by injecting the hard input mass scale,  $M$, into the theory.\footnote{One could
allow $\Lambda_{QCD}$
to define $\alpha_3(\mu)$
if one is only considering QCD, but this is otherwise a ``bottoms up'' view
of the world in which, \eg, grand unification makes no sense. Here we assume a ``top-down'' 
understanding where low energy physics is determined by ultra-high energy
dynamics at $M>>\Lambda_{QCD}$. This implies the notion of ``dimensional transmutation,''
whereby a dimensionless coupling constant, $\alpha_3(\mu)$,
defines a mass-scale $\Lambda_{QCD}$ with no other
inputs, is not applicable. $\alpha_3(\mu)$ will always require
some high energy boundary condition $\alpha_3(M)$ with implicit scale $M$.}

To maintain the scale symmetry, $M$ must be replaced by something
in the action, such as the VEV of some high energy
field, $\Phi$, such as the $\{\bf 24\}$
in  $SU(5)$, or 
some other dynamical scale. As we'll see in our effective
potential calculation the most natural choice of $\Phi$ is $ \sim \sqrt{K}$,
the order parameter.
The RG of QCD generates a large heirarchy, but
the strong scale $\Lambda_{QCD}$ is determined by the particular RG trajectory defined
by $\alpha_3(\Phi)$, and thus $\Lambda_{QCD}\propto \Phi$. 
The scale symmetry thus remains intact in
eq.(\ref{QCD}) since, holding the numerical input value $\alpha_3(\Phi)$ fixed,
 $\Lambda_{QCD}$ and $\Phi$ are rescaled  together and the ratio
is fixed. 

This illustrates the general result that in
Weyl invariant theories there are no hard input mass scales, and
mass is only defined by ratios of VEV's.
The QCD scale is inherited from the scale $\Phi$, and not dimensionally
transmuted out of nothing: ``dimensional inheritance'' 
may be a better moniker for
this phenomenon.

If no additional mass scales, beyond of the
dynamical VEV's of the fields present in the action are introduced into the
final renormalized action, then Weyl symmetry can remain intact. This
is contrary to the usual method of computation:  usually an external 
mass scale $M$ is introduced when a regularized amplitude, \eg, 
such as a Coleman-Weinberg potential \cite{CW},
is renormalized.  This leads directly to the Weyl current
anomaly (or scale current trace anomaly) as we  discuss in V.A. 
If instead, one replaces $M$ by
$\sqrt{K}$, 
then the anomaly is cancelled, and the Weyl symmetry is
maintained.  The potential is then a function
of fields only and contains dimensionless ratios of fields as
the arguments of logs.
We will demonstrate how this works with an 
explicit calculation of 
a scale symmetric Coleman-Weinberg potential in V.B. 
This will also demonstrate the inertial breaking of a ``flavor'' $U(1)$
symmetry.

We begin with a brief overview of Weyl symmetry in Section II. While
we mainly focus on globally Weyl invariant theories,
we sketch how the
approach can be implemented in both global and local Weyl
invariant theories, where the latter introduces Weyl's ``photon''
$\hat{A}_\mu$. We then discuss the phenomenon
of inertial symmetry breaking in Sections III and IV. Finally we turn to the 
effects of quantum
mechanics in Section V and conclusions in Section VI.

\section{Weyl Symmetry in a Nutshell}

Many years ago Hermann Weyl had the idea that, since coordinates are merely
numbers invented by humans to account for events in space-time, they
should not carry length scale \cite{Weyl}.  Rather, the concept  of length
should be relegated
to the (covariant) metric, and (contravariant) coordinate differentials are scale free.
Therefore, under a local Weyl scale transformation we would have:
\bea
g_{\mu\nu}(x) \rightarrow e^{-2\epsilon(x)}g_{\mu\nu}(x) \qquad
&& g^{\mu\nu}(x)\rightarrow e^{2\epsilon(x)}g^{\mu\nu}(x) \nonumber \\
\sqrt{-g}\rightarrow e^{-4\epsilon(x)}\sqrt{-g}\qquad 
&&\phi(x)\rightarrow e^{\epsilon(x)}\phi(x)
\eea
Here we've included the local transformation of a scalar field which transforms
like a mass (length)$^{-1}$. The contravariant metric must 
transform as (length)$^{-2}$ to preserve the condition: 
$g_{\mu\nu}g^{\nu\rho}=\delta_\mu^\rho$.
Weyl transformations are distinct from coordinate diffeomorphisms that 
define scale transformations on coordinates,
as $\delta x^\mu=\epsilon(x) x^\mu$, which we discuss below.  
The global Weyl symmetry 
corresponds as usual to $\epsilon=$(constant in spacetime).  

It is straightforward to construct a list of local Weyl invariants:
\bea
\!\! \!\!\phi^2(x)g_{\mu\nu}(x); \!\! \! && \phi^{-2}(x)g^{\mu\nu}(x); %\nonumber
\;\;\;
\sqrt{-g}(x)\phi^4(x); \nonumber 
\\
R(\phi^2g_{\mu\nu})&=&\phi^{-2}R(g_{\mu\nu})+6\phi^{-3}D^\mu(\partial_\mu\phi) \nonumber
\\
\label{relation}
\sqrt{-g}\phi^4R(\phi^2g_{\mu\nu})&=&\sqrt{-g}\phi^2R(g_{\mu\nu})
+6\phi D^\mu(\partial_\mu\phi)%\nonumber 
\\ ...\nonumber
\eea
Note that the computation of $R(\phi^2g_{\mu\nu}) $ above
requires that any Christoffel symbols used in the definition of $R$ be evaluated 
in the metric $\phi^2 g_{\mu\nu}$.
Using these identities we can construct an action that is locally
Weyl invariant:
\bea
S&=&\int d^4x\;\sqrt{-g}\left(-\frac{1}{12}\phi^4R(\phi^2g)-\frac{\lambda}{4}\phi^4 \right)\nonumber \\
&=&\int d^4x\;\sqrt{-g}
\left(\frac{1}{2}\partial_\mu\phi\partial^\mu\phi-\frac{1}{12}\phi^2 R(g)-\frac{\lambda}{4}\phi^4 \right)  
%\nonumber 
\eea
where we substituted the relationship of eq.(\ref{relation}) and integrated by parts using the divergence
rule $D_\mu V^\mu= \sqrt{-g}^{-1}\partial_\mu (\sqrt{-g}V^\mu)$. Here we obtain the
famous locally Weyl invariant theory in which the nonminimal coupling of scalars to gravity
is fixed by the coefficient $1/12$, needed to canonically normalize the
$\phi$ kinetic term.  This is a special and somewhat degenerate
theory, since we
can revert to the metric $\hat{g}_{\mu\nu}=\phi^2 g_{\mu\nu}$ and $\phi$ disappears
from the action.
The theory has a vanishing Weyl current \cite{JP}.

We note that covariant gauge fields, 
such as the electromagnetic vector potential, $A_\mu $,
do not transform under the local Weyl transformation,  since they are associated with 
 derivatives $\partial_\mu-ie A_\mu $ which, like coordinates, do not transform. The 
 electromagnetic fields that have the usual engineering scale 
 $\sim$(mass)$^2$, $\vec{E}$ and $\vec{B}$, are contained in
 the field strength with one covariant and one contravariant index,
 $F_\mu^{\;\;\nu}$, \eg, $\vec{E}_i = F_i^0$.

We can construct
a covariant derivative of a scalar field under local Weyl transformations by
introducing the ``Weyl photon,''  $\tilde{A}_\mu$, as 
\beq
\tilde D_\mu\phi=\partial_\mu\phi -\tilde{A}_\mu\phi
\eeq
where $\phi(x)\rightarrow e^{\epsilon(x)} \phi(x)$ and 
$\tilde{A}_\mu(x)\rightarrow \tilde{A}_\mu(x) +\partial_\mu \epsilon(x)$  
(note the major difference from electrodynamics in the absence of a factor of $i$
in the coefficient of $\tilde{A}_\mu$).
%We will postpone discussion of
%this construction momentarily, and focus presently on global Weyl tranormations for which
%$\epsilon$ is constant and $\partial_\mu \phi\rightarrow e^\epsilon\partial_\mu \phi$
%is covariant.
Armed with this we can construct another local Weyl invariant:\footnote{  
Here there is a subtlety. We must define the derivative of any
conformal field as a commutator: $[D_\mu, \Phi] =\partial_\mu\Phi -A_\mu[W,\Phi]$
where $[W,\phi] = w\phi$ and $w$ is the conformal charge of $\Phi$.
Hence $w=1$ for $\phi$. We also require $w=-2$ for $g_{\mu\nu}$,
$w=+2$ for $g^{\mu\nu}$, $w=-4$ for $\det{-g}$, etc. Note that
$[D_\mu, g_{\rho\sigma}] =D_\mu g_{\rho\sigma} +2\tilde A_\mu g_{\rho\sigma}$
$=\tilde A_\mu g_{\rho\sigma}$ since $D_\mu g_{\rho\sigma}=0$.
This insures
the invariance of the action with the Weyl covariant derivative 
under integration by parts.
Note that we can alternatively define a restricted
``pure gauge theory'' with $A_\mu =\partial_\mu \ln(\chi)$,
where $\chi$
is any massless scalar field.}
\beq
\sqrt{-g} g^{\mu\nu} \tilde D_\mu\phi(x) \tilde D_\nu\phi(x).
\eeq
This is a locally invariant kinetic term. We can combine it 
with the previous invariants to define an action in which the
Weyl symmetry is local, yet the
nonminimal coupling of scalars to $R$ is arbitrary:
\bea
\label{local}
S&=&\int d^4x\;\sqrt{-g}\left(
 \half (1-\alpha) g^{\mu\nu}\tilde D_\mu\phi \tilde D_\nu\phi
-\frac{\lambda }{4}\phi^4 \right. \nonumber \\
& &\qquad  \qquad \qquad \left.
-\frac{\alpha}{12}\phi^4R(\phi^2g)\right)\nonumber \\
&=&\int d^4x\;\sqrt{-g}
\left(
\frac{1}{2}g^{\mu\nu}\partial_\mu\phi\partial_\nu\phi-\frac{\alpha}{12}\phi^2 R(g)
-\frac{\lambda}{4}\phi^4
\right.
\nonumber \\
&& \qquad \left.
- \half (1-\alpha)\left(\tilde{A}^\mu  \partial_\mu(\phi^2) -
\tilde{A}^\mu\tilde{A}_\mu \phi^2 \right) \right)  .
\eea
From the action of eq.(\ref{local}) we see that the Weyl current is:
\bea
K_\mu & =& -\frac{1}{\sqrt{-g}}\frac{\delta S}{\delta \tilde A^\mu}
%\left|_{}_{\tilde A_\mu =0}
= (1-\alpha)\left( \phi\partial_\mu\phi-\tilde{A}_\mu\phi^2\right)
\nonumber \\
&=& (1-\alpha)\phi\tilde{D}_\mu\phi .
\eea
By setting $\tilde{A}_\mu=0$ we obtain
a globally invariant theory, and this current
becomes the conserved Noether current for
the global Weyl invariant theory:
\beq
\label{current}
K_\mu 
= (1-\alpha) \phi\partial_\mu\phi.
\eeq
As stated in the introduction, we then have:
\beq
K_\mu = \partial_\mu K ; \;\;\qquad K = \half (1-\alpha) \phi^2
\qquad 
\eeq
The structure
and conservation law of
$K_\mu$ also follows directly from the Einstein and Klein-Gordon
equations in the Jordan frame \cite{FHR}, which we 
demonstrate in Section III. 

We further remark that,
in the local Weyl photon theory, we can include a kinetic term:
$(-1/4g^2)\hat{F}_{\mu\nu}\hat{F}^{\mu\nu}$ where 
$\hat{F}_{\mu\nu}=\partial_{\mu} \hat{A}_{\nu }-\partial_{\nu} \hat{A}_{\mu }$.
When the theory inertially breaks, the dilaton will be eaten by
$\hat{A}_\mu$, giving a massive Weyl photon, of mass $ 2g\sqrt{\bar{K}}$.

How does scale symmetry differ from Weyl symmetry?
Scale symmetry is generally codified as a diffeomorphism.
In a flat spacetime, where we hold the metric fixed $g_{\mu\nu}=\eta_{\mu\nu}$,
the coordinates and fields transform under a local Noether variation 
$\epsilon(x)$ as:
\bea
\delta x^\mu &= &\epsilon(x)\;x^\mu \nonumber \\
\delta \phi(x) &= &-\epsilon(x)\phi(x)+\epsilon(x) x^\mu\partial_\mu \phi(x).
\eea
 If a theory with
action $S$ is scale
invariant, then we find $J^\mu = -\delta S/\delta\partial_\mu\epsilon=x^\nu T^{\mu}_{\nu}$
is the scale current,
where   $T^\mu_\nu$ is the canonical stress tensor. Taking the divergence 
we immediately have $\partial_\mu J^\mu =  T^\mu_\mu$, which is the trace
of stress tensor. 

However, there are traditional difficulties in 
defining the scale current in the case of a formally massless scalar field, $\phi$.
Even in a scale invariant classical
theory we obtain $T^\mu_\mu\neq 0$, unless we construct the ``improved stress tensor,''
$T^I_{\mu\nu}$, of
Callan-Coleman-Jackiw (CCJ) \cite{CCJ}.  The improved stress tensor can be obtained
by allowing an arbitrary metric, $g_{\mu\nu}$, with the addition
of a nonminimal coupling of $\phi$ to curvature, $-(1/12)\phi^2R$. We then
compute $T^{\mu\nu}$ by variation of the action wrt $g_{\mu\nu}$,  
and then impose the flat space limit, $g_{\mu\nu}=\eta_{\mu\nu}$. 
The result is the improved
stress tensor of CCJ, and corresponds to the {\em rhs} of 
our eq.(\ref{einstein1}) below, with $\alpha=1$: 
\bea 
\label{imp1}
T^{I}_{\mu\nu} & = &
%\left(
\frac{2}{3} 
%\right)
(\partial_{\mu }\phi \partial_{\nu }\phi)
-\frac{1}{6} (g_{\mu \nu }
%\left( 
%\right) 
\partial^{\lambda }\phi \partial_{\lambda }\phi )
\nonumber \\
& & \!\!\!\!\!  \!\!\!\!\! \!\!\!\!\!    
+\frac{1}{3} \left( g_{\mu \nu}\phi \partial^{2}\phi -\phi \partial_{\mu }\partial_{\nu }\phi \right) 
+g_{\mu \nu}V(\phi). 
\eea
With the scale current $J_\mu =x^\nu T^{I}_{\mu\nu} $, 
the trace is the {\em rhs} of eq.(\ref{Kdiv1}) below, with $\alpha=1$:
\bea 
\label{imp2}
T^{I}{}_{\mu}^{\mu}& = & \phi D^{2}\phi +4V(\phi)= -\phi \delta V/\delta \phi +4V(\phi).
\eea
For a scale invariant theory, $V(\phi)\propto \phi^4$,
hence $\phi \delta V/\delta \phi = 4V$. The trace then vanishes 
by the Klein-Gordon equation, eq.(\ref{KG1}) below (see also \cite{CTH}).

Hence, the improved stress tensor is
that of a Weyl invariant theory in the flat space limit.
Not surprisingly, the trace is now
identically the divergence of the Weyl current.
If Weyl (scale) symmetry is conserved (broken) then
scale (Weyl) symmetry will be conserved (broken).
For practical purposes, Weyl symmetry of a theory in $D=4$ contains as much
information as the diffeomorphism scale symmetry.
The Weyl symmetry is actually more
convenient to implement than scale diffeomorphisms since it  does not
involve shifting coodinate arguments of fields. 

%\newpage

\section{Tale of Two Actions}

Let us now see how the Weyl current emerges and
plays a central role by a simple example.
Consider a real scalar field theory action together with 
Einstein gravity and a cosmological constant
(our metric signature convention is $(1,-1,-1,-1)$):
\beq
\label{eqone}
S=\int \sqrt{-g}\left(
\frac{1}{2}g^{\mu \nu }\partial_{\mu }\sigma \partial_{\nu}\sigma 
  -\Lambda +\half M_P^2 R \right)  .
\eeq
This action provides a caricature of the cosmological world we live in. 

We imagine an initial, ultra-high-temperature phase in which the massless
scalar $\sigma$ has the dominant energy density, $\rho_\sigma \propto T^4$. 
Consider a Friedman-Robertson-Walker (FRW) metric:
\bea
g_{\mu\nu }& =&  [1,-a^{2}(t),-a^{2}(t),-a^{2}(t)] \qquad H=\frac{\dot{a}}{a} \;.
\eea 
In this theory the universe initially 
expands in a FRW phase, with the temperature
red-shifting as $T\sim 1/a(t)$, and the scale factor growing as $a(t)
\sim \sqrt{t}$.  Eventually the $\sigma$ thermal energy becomes smaller
than the cosmological constant, $\rho_\sigma < \Lambda$, and we then enter a deSitter
phase with exponential growth, $a(t)\sim e^{t\sqrt{\Lambda/3M^2_{P}}}$.
We can model the thermal phase as a pre-inflationary era, and the 
cosmological constant then represents a potential energy that drives inflation.
In any case, the intuition that allows us to readily understand how this works 
is well-honed.

Now consider a  different  action:
\beq
\label{eqtwo}
S=\int \sqrt{-g}\left(\frac{1}{2}g^{\mu \nu }\partial_{\mu }\phi \partial _{\nu
}\phi  -\frac{\lambda}{4}\phi^4 -\frac{\alpha}{12}\phi^2 R \right)  .
\eeq
This action is scale invariant, having no cosmological constant or
Planck scale.  In fact, this is precisely the action of
eq.(\ref{local}) with $\tilde{A}_\mu=0$ and is globally Weyl invariant.

It turns out, as we see below, that these two theories 
are mathematically equivalent, provided $\alpha<1$.
Our question then becomes, given that, from our accumulated experience in 
inflationary cosmology we understand the dynamics of
eq.(\ref{eqone}) so well,  then how
can we easily understand the dynamics of eq.(\ref{eqtwo})?
At first this doesn't look too hard.  Indeed, if $\phi$ starts out in 
some very high-temperature phase, where the energy density
is large compared to $\lambda \phi^4$
then we expect the scale factor will increase
in a scale invariant way, $a(t)\sim t$. This follows by intuiting that the Hubble constant
satisfies $H^2\sim T^4/\phi^2$, where the $\phi^2$
factor in the denominator replaces $M_P^2$. In thermal equilibrium we expect $\phi^2 \sim T^2$ 
and thus $H=\frac{\dot{a}}{a}\sim T\sim \frac{1}{t}$. Therefore, $a(t)\sim t$.

However, as the universe cools, we expect ${\phi}(x)$ to settle into some spatially
constant VEV  $\VEV{\phi}$.  Our intuition from conventional $M_P^2R$ gravity tells us that
this VEV will slow-roll in the potential, with $\VEV{\phi}$ eventually becoming zero. 
However, in eq.(\ref{eqtwo})
this would imply a vanishing $M_P$,   and the details of the
solution are less clear.  It is plausible that the increasing strength of gravity
will increase the Hubble damping, and halt the relaxation of $\VEV{\phi}$,
perhaps leading  to a nonzero cosmological constant $\lambda \VEV{\phi}^4$. 
If true, this would then match the 
cosmological constant case of eq.(\ref{eqone}), and it would imply a spontaneous
breaking of scale symmetry.  But how can we see that this happens in 
a simple and intuitive way, without having to puzzle over the solutions
of coupled nonlinear differential equation?

Indeed, from eq.(\ref{eqtwo}) we can directly obtain the Einstein equation:
\bea 
\label{einstein1}
 \!\!\!
\frac{1}{6}\alpha \phi ^{2}G_{\alpha \beta } & = &
\left(\frac{3-\alpha}{3} \right)\partial _{\alpha }\phi \partial _{\beta }\phi
-g_{\alpha \beta }\left( \frac{3-2\alpha}{6}\right) \partial
^{\mu }\phi \partial _{\mu }\phi 
\nonumber \\
& & \!\!\!\!\!  \!\!\!\!\! \!\!\!\!\! \!\!\!\!\!     
+\frac{1}{3}\alpha \left( g_{\alpha \beta
}\phi D^{2}\phi -\phi D_{\beta }D_{\alpha }\phi \right) +g_{\alpha \beta
}V(\phi) .
\eea
The trace of the Einstein equation becomes:
\bea 
\label{trace1}
\!\!\!\!\!\!\!\!\!\! \!\!\!\!\!\!
-\frac{1}{6}\alpha \phi ^{2}R & = &  (\alpha-1 )\partial ^{\mu }\phi
\partial _{\mu }\phi +\alpha \phi D^{2}\phi +4V(\phi).
\eea
We also have the Klein-Gordon (KG) equation for $\phi$:
\bea
\label{KG1}
0=\phi D^{2}\phi+\phi\frac{\delta }{\delta \phi }V\left( \phi
\right) +\frac{1}{6}\alpha \phi^2 R.
\eea
We can combine the KG equation, eq.(\ref{KG1}), and trace equation, 
eq.(\ref{trace1}), to eliminate the $ \alpha \phi^2 R $ term, and obtain:
\bea
\label{Kdiv00}
0& =& (1-\alpha )\phi D^{2}\phi +(1-\alpha )\partial ^{\mu }\phi \partial
_{\mu }\phi 
\nonumber \\
& & \qquad\qquad +\phi \frac{\delta }{\delta \phi }V\left( \phi \right) -4V\left(
\phi \right) .
\eea
This can be written as a current divergence equation:
\bea
\label{Kdiv1}
D^\mu K_\mu = 4V\left(
\phi \right) -\phi \frac{\partial }{\partial \phi }V\left( \phi \right) .
\eea
where $K_\mu=(1-\alpha)\phi\partial_\mu \phi$ is the Weyl current as given in eq.(\ref{current}).
For the scale invariant potential, $V(\phi)\propto \phi^4$, the {\em rhs}
of eq.(\ref{Kdiv1}) vanishes
and the $K_\mu$ current is then covariantly conserved:
\beq
\label{cons1}
D^\mu K_\mu = 0.
\eeq
We see that this is an ``on-shell'' conservation law, \ie, 
it assumes that the gravity satisfies eq.(\ref{einstein1}).

We can now understand the behavior of
this theory by the spontaneous breaking of the Weyl symmetry. 
Starting with an arbitrary classical $\phi$, after a period
of general expansion, in some arbitrary patch of space, 
 $\phi$ becomes approximately spatially constant, but time dependent,
hence, $D^2 \phi =\overset{..}\phi+3H\dot{\phi} $.
The Weyl current conservation law of eq.(\ref{cons1}) then becomes:
\bea
\label{cons2}
\ddot K+3H \dot K=0.
\eea
If we take $\phi$ to be a function of time $t$ only, 
we have by eq.(\ref{cons2})
\beq
K(t)=c_1 + c_2\int_{t_0}^t \frac{dt'}{a^3(t')}
\eeq
where $c_1$ and $c_2$ are constants which are determined
by initial values of $(\phi, \dot\phi)$. Therefore, under arbitrary
initial conditions:
\beq
 K(t\rightarrow\infty)\rightarrow\bar{K} \;\;\makebox{constant,}
\eeq
and it follows that: 
\beq
\phi(t\rightarrow\infty)\equiv\phi_0,\;\; \makebox{hence}, \;\;
\bar{K}=\frac{1}{2}(1-\alpha)\phi_0^2.
\eeq
$\bar{K}$ is the constant asymptotic value 
of the order parameter, and defines the broken Weyl symmetry
phase.

The Planck mass and cosmological constant are then determined:
\beq
M_{P}^2 = -\frac{\alpha}{3(1-\alpha)}\bar{K},
\qquad \Lambda= \frac{\lambda \bar{K}^2}{(1-\alpha)^2}.
\eeq
Then the $(00)$ Einstein equation, with $G_{00}=-3H^2$, gives
\beq
\label{fixed1}
 H^2=-\frac{\lambda\phi_0 ^{2} }{2\alpha }=-\frac{\lambda\bar{K} }{\alpha(1-\alpha) }
 =\frac{\Lambda}{3M_{P}^2}.
\eeq
This gives eternal inflation if
$\alpha <1$ with constant $K=\bar{K}$, which
matches the Einstein frame conclusions.

We have thus understood how the behavior of eq.(\ref{eqtwo})
matches that of eq.(\ref{eqone}) by way of the $K$-current! 
We never had to solve the complicated eqs.(\ref{einstein1},\ref{KG1}),
as the behavior is dictated by the ``current algebra.'' We have also
gained insight into how the Weyl symmetry is broken, as the Jordan frame
theory of eq.(\ref{eqtwo}) flows into the Einstein frame of eq.(\ref{eqone})
under the general pre-inflationary expansion, and the relaxation of $K_0\rightarrow 0$,
and $K\rightarrow \bar{K}$.

The asymptotic field $\phi(t\rightarrow\infty)\equiv\phi_0$ is a particular solution, 
and is subject to ``small fluctuations.''
In eq.(\ref{eqtwo}) we identify $\phi$ with
the asymptotically constant field VEV, $\phi_0 $, and include a 
small fluctuation field, $\sigma /f$,
where $f=\sqrt{2\bar{K}}$,  as:
\beq
% \exp (-\sigma/f )\phi =\phi _{0}\qquad 
\phi =\phi _{0}\exp (\sigma/f ),
\eeq
and perform the Weyl metric transformation: 
\beq
g_{\mu \nu }=\exp (-2\sigma/f )\widetilde{g}_{\mu \nu },
\eeq
we obtain \cite{FHR}: 
\bea
\label{fixed2}
S=\int \sqrt{-\widetilde{g}}\left( \frac{1}{2}\widetilde{g}^{\mu \nu
}\partial _{\mu }{\sigma } \partial _{\nu }{\sigma }
-\Lambda +\frac{1}{2} M_P^{2} R(\widetilde{g})
\right) .
\eea 
Therefore, ``small deformations'' of $\phi_0$ 
are described by $\sigma$.  We also
see that the scale invariant theory, 
 eq.(\ref{eqtwo}), defined in the ``Jordan frame''
is equivalent to the ``Einstein frame'' theory eq.(\ref{eqone},\ref{fixed2})
in the broken symmetry phase, where we have traced the origin of the
mass scales to the inertial spontaneous symmetry breaking order
parameter $K$. 

The massless  field 
${\sigma }$ is, of course, the dilaton, with ``decay constant'' $f$. 
The variation of the action of eq.(\ref{fixed2})
with respect to $\sigma/f$
yields the form of the current in the broken phase,  $K_\mu = f\partial_\mu\sigma $,
the  analogue of the pion axial current, $f_\pi \partial_\mu \pi$.
The dilaton reflects the fact that the
exact underlying Weyl symmetry remains intact, though it is hidden in the Einstein frame.
We can rescale both the VEV $\phi_0 \rightarrow e^{\epsilon} \phi_0$
and the Hubble constant $H_0\rightarrow e^{\epsilon} H_0$ while their ratio remains fixed:
\beq
 \frac{ H_0^{2}}{\phi _{0}^{2}} = \frac{\lambda }{2\left| \alpha \right| }.
\eeq 
It is straightforward to extend this to include matter fields \cite{FHR2}.
If all ordinary masses arise only from the spontaneous breaking of the 
Weyl symmetry, then the dilaton only couples directly to gravity. There are then
no Brans-Dicke-like constraints,  \cite{FHR2}.

\section{Two Scalar Theory}

We now consider a more realistic
 $N=2$ model, with scalars $(\phi, \chi)$, and the potential:
\bea
\label{pot2}
W\left( \phi ,\chi \right) =\frac{\lambda}{4} \phi ^{4}+\frac{\xi}{4} \chi ^{4}+\frac{\delta}{2} \phi
^{2}\chi ^{2}.
\eea
The action takes the form:
\bea
\label{action2}
S &= & \int \sqrt{-g}\left( \frac{1}{2}g^{\mu \upsilon }\partial _{\mu }\phi
\partial _{\nu }\phi +\frac{1}{2}g^{\mu \upsilon }\partial _{\mu }\chi
\partial _{\nu }\chi
\right.
\nonumber \\
& & \left.
 -W\left( \phi ,\chi \right) -\frac{1}{12}\alpha_1 \phi^{2}R
 -\frac{1}{12}\alpha_2 \chi ^{2}R\right) .
\eea
This has been studied extensively in \cite{shapo,GB,FHR}.

We presently follow the approach of \cite{FHR} and work directly
in the defining Jordan frame of eq.(\ref{action2}), 
and then just follow the dynamics.  
The sequence of steps follows those of the previous
single scalar case. 
The trace of the Einstein equation immediately becomes:
\bea
M_P^2R & = & \left(
(\alpha_1-1 )\partial ^{\mu }\phi\partial _{\mu }\phi+\left(\alpha_2-1 \right) \partial ^{\mu }\chi \partial _{\mu }\chi 
\right.
\nonumber \\
& & \left.
+\alpha_1 \phi D^{2}\phi +\alpha_2 \chi D^{2}\chi +4W\left(
\phi ,\chi \right) \right).
\eea
where,
\bea
M_P^2=-\frac{1}{6}\left( \alpha_1 \phi ^{2}+\alpha_2 \chi ^{2}\right).
\eea
The Klein-Gordon equations for the scalars are:
\bea
0 & =& D^{2}\phi+
\delta \phi\chi^{2}+\lambda \phi^{3}+\frac{1}{6}\alpha_1\phi R
\nonumber \\
0 & =& D^{2}\chi+
\delta \phi^{2}\chi+\xi \chi^{3}+\frac{1}{6}\alpha_2\chi R
\eea
and we again use the trace equation 
and the Klein-Gordon equations to elimate $R$.
We obtain:
\bea
D_{\mu }K^{\mu } & =&  4W-\phi\frac{\delta W}{\delta\phi}-\chi\frac{\delta W}{\delta\chi},
\eea
where now:
\beq
\label{Kcurrent2}
K_{\mu }=(1-\alpha_1 )\phi \partial_{\mu }\phi +\left( 1-\alpha_2 \right) \chi
\partial_{\mu }\chi .
\eeq
The current can be written in terms of a kernel as $K_{\mu }=\partial_\mu K$ 
and is conserved for a scale invariant potential, $W(\phi,\chi)$.

\begin{figure}[tbp]
%\centering
\vskip-0.1in
\begin{flushleft}
\includegraphics[width=8.9cm]{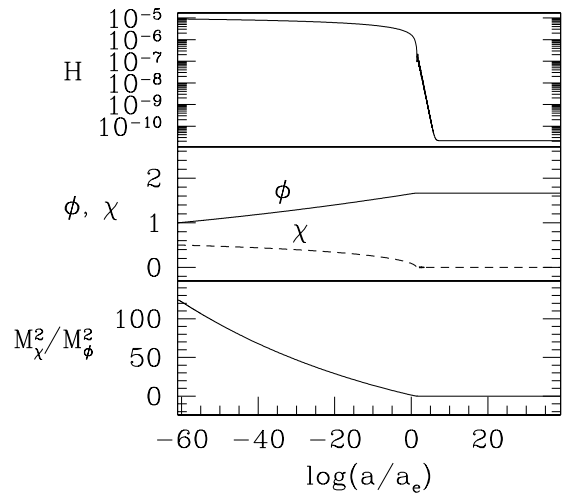} 
\end{flushleft}
\vspace{-3cm}
\caption{{Plot of the Hubble parameter, $H$, $\phi$, $\chi$ and the 
ratio of the two components of the effective Planck mass, $M^2_\phi$ and $M^2_\chi$, 
as a function of $a$; we have normalized the x-axis to the scale factor at the end of inflation, $a_e$. }}
\label{figure_back}
\end{figure}

This leads to a realistic cosmological evolution as illustrated 
in Fig.\ref{figure_back}.
During an initial pre-inflationary phase, a small patch of the universe will redshift from arbitrary initial 
field values and velocities,  $(\phi, \dot{\phi}; \chi,\dot{\chi})$. 
The redshifting will 
cause $(\dot{\phi},\dot{\chi})\rightarrow 0$ and the $K_0$ charge density will dilute away as $\sim a(t)^{-3}$.
As in the simple single scalar case this leads to   a state with constant kernel $K\rightarrow \bar{K}$.
which again determines the Planck mass, $\bar{K}\sim M_P^2$, and
 spontaneously breaks scale symmetry. 
 
When a constant $K\approx \bar{K}$ is attained,
the fields VEV's $(\VEV{\phi},  \VEV{\chi)}$
are constrained to lie
on an ellipse defined by:
\beq
\bar{K}=\half\left[(1-\alpha_1 )\VEV{\phi}^2 +\left( 1-\alpha_2 \right) 
\VEV{\chi}^2\right].
\label{Kernel2}
\eeq
(this ellipse condition is, to our knowledge, first discussed in \cite{GB}).
The pre-expansion has caused the fields to fall to approximate initial locations
on the ellipse, $(\VEV{\phi}_0,  \VEV{\chi}_0)$,
which are random. This trapping of the fields on the ellipse leads to
inflation.

We now make an additional assumption
that the potential of eq.(\ref{pot2}) has a flat direction.
For example, we can take the specific form:
\bea
\label{pot3}
W\left( \phi ,\chi \right) =\frac{\lambda}{4}\left(\chi^2 -\eta \phi ^{2}\right)^2.
\label{pot1}
\eea
The flat direction $\chi=\sqrt{\eta} \phi$  defines a  ray in the $(\phi,\chi)$ plane that intersects the 
ellipse.  
The random initial values of $(\VEV{\phi}_0,\VEV{\chi}_0)$ on the ellipse
would generally not
be expected to coincide with the flat direction. 

For a significant region of the values on the ellipse, the fields then slow-roll, 
migrating toward the intersection 
of the potential and generating a finite period of inflation.  
If we assume $\eta<<1$ this intersection occurs
near the right-most  end of the ellipse where $\VEV{\chi} << \VEV{\phi}$
in the quadrant, $(\VEV{\phi},\VEV{\chi})>0$.
The inflationary period ultimately ends as the fields approach the intersection
of the flat direction with the ellipse. When the slow-roll conditions are violated  
and the system enters a period of ``reheating,'' the potential energy 
is converted to kinetic energy which rapidly redshifts.

This theory generates random trapped values on the ellipse
and has no serious initial condition
fine tuning problem. Here inflation is not governed by dwelling
at a maximum in a potential, but rather by the inertial scale breaking
and the initial period of formation of the value
$\bar{K}$, and subsequent 
slow migration along the ellipse.
It is easy to arrange parameters to establish adequate inflation and 
generate observables consistent with cosmological data.
Our most recent analysis of cosmological observables and comparisons to the
parameters of the model are given in ref.\cite{FHRN}.

\section{Quantum Theory}

\subsection{Weyl Invariant Effective Potentials}

For the scenario of inertial spontaneously broken Weyl symmetry to work,
and lead to a stable Planck mass,
it is essential that the Weyl current be identically conserved:
\beq
\label{cons3}
D_\mu K^\mu = 0.
\eeq
Failure of this is not an option! In what follows we will refer
to nonzero contributions coming from loops to the rhs of eq.(\ref{cons3}) as 
``Weyl anomalies.''  As we've seen in Section II, the trace anomalies of
the scale current defined by diffeomorphisms are identical to those of $K_\mu$.

Scale and Weyl symmetry of a theory appears ab initio to be broken by 
arbitrary divergences of quantum loops.
Loop divergences are subtle, however,  and are often confused with physics.
Here we adopt an operating principle that has been espoused by W. Bardeen \cite{bardeen}:
The allowed symmetries of a 
renormalized quantum field theory are
determined by anomalies, (or absence thereof). Other quantum loop divergences are 
essentially unphysical artifacts of the 
method of calculation. 

Our problem  of maintaining Weyl symmetry
requires that we build a theory that has no anomaly in $K_\mu$.
To understand this problem, and its solution,
we turn the the Coleman-Weinberg potential \cite{CW}.  

In computing Coleman-Weinberg potentials for massless scalar fields
 we encounter an infrared divergence that must be
regularized. To do so  we often
introduce explicit ``external'' mass scales into the theory by hand. These are mass scales
that are not part of the defining action of the theory, and essentially
define the RG trajectories of coupling constants.
These externally injected mass scales  lead  directly to the Weyl anomaly.

 We can see this in
the famous paper of Coleman and Weinberg \cite{CW}. In their eq.(3.7), to {\em renormalize}
the quartic scalar coupling constant in an effective potential at one-loop level,
$V(\phi)$, they introduce a mass scale $M$.  Once one injects $M$ into the theory,
scale and Weyl symmetries are broken, and the
effective potential in the leading $\ln(\phi/M)$ limit then takes the form: 
\beq
V(\phi)=\frac{\beta_1}{4} \phi^4 \ln(\phi/M).
\eeq
Here $\beta_1 $ is the one-loop renormalization group $\beta$-function, 
$ {d}\lambda(\mu)/{d\ln\mu}=\beta_1$.   We emphasize that this is not an issue
of regularization, but rather stems from the process of renormalization.

The manifestation of this 
is seen in the Weyl anomaly, (and trace of the improved stress tensor \cite{CCJ}),
\ie, the divergence of the $K_\mu$ current:
\beq
\label{traceanom}
\partial_\mu K^{\mu} =4V(\phi)-\phi\frac{\partial}{\partial\phi}V(\phi)
=-\frac{\beta}{4} \phi^4.
\eeq
The anomaly is 
directly associated with the $\beta$-function
of the potential coupling constant $\lambda$.  The coupling
constant $\lambda(\phi)$ is viewed to run with the field (VEV)
as determined by the RG equation.
A specification
of the boundary condition of the RG trajectory has injected
$M$ into the theory, 
 \eg, $\lambda(\phi = M)=0$).
We clearly see in  eq.(\ref{traceanom})  that the anomaly 
is then generated. This is similar to
the QCD case described in Section I.

Of course, depending upon the application,
there's nothing wrong with the Coleman-Weinberg
potential defined this way. If one is only treating the effective
potential as a subsector of the larger theory, where $M$ is 
truly external to the subsector,
then we can simply defer the question 
of the true origin of $M$ in the larger theory.
%If, however,  we are interested in the more general question, {\eg}, of  the overall
%scale behavior 
%of a ``theory of everything,'' 
%we have to understand the  origin of $M$.  
If, however, Weyl symmetry is to be maintained as an exact invariance
of the overall theory, then  $M$ {\em must be replaced by an internal 
mass scale that is part of action, }
\ie, $M$ must then be (the VEV of)
 a dynamical entity, suxh as a field $\chi$ (or $K$) appearing in the extended action. 
We would then have the Coleman-Weinberg potential:
\beq
V(\phi,\chi)=\frac{\beta_1}{4} \phi^4 \ln({\phi}/{\chi}).
\eeq
(or $\chi \rightarrow c\sqrt{K}$) and then we have the Weyl anomaly:
\beq
\label{traceanom2}
\partial_\mu K^{\mu} =\left(4-\phi\frac{\partial}{\partial\phi}
-\phi\frac{\partial}{\partial\chi}\right)V(\phi,\chi) =0.
\eeq
The $\beta -$function contributions have now cancelled. 
To see how the argument of the log might be specified
more precisely we turn to an explicit calculation
in the next section.

\subsection{$U(1)$ Model}

Inertial symmetry breaking yields a spontaneously generated mass scale which
becomes the Planck mass and  can produces other scales
in the theory. It can also drive the breaking of other
symmetries.  
To see this, 
let us revisit the two scalar theory with the nonminimal coupling
to gravity of the form:
\bea
\label{oneU1}
S&=&\int \sqrt{-g} g^{\mu \upsilon }\left(\frac{1}{2}\partial _{\mu }\phi_1
\partial _{\nu }\phi_1 +\frac{1}{2}
\partial _{\mu }\phi_2\partial _{\nu }\phi_2\right)
\nonumber \\
&&
-\int d^4x\sqrt{-g}\left(\frac{\alpha}{12}\left(\phi_1^2 + \phi_2^2\right)R
+W \right)
\eea
where  $W=W(\phi_1,\phi_2)$.
We now have a common overall value of $\alpha$.  Apart from
the potential, this theory
thus has a $U(1)=SO(2)$ symmetry, and we can introduce
a complex field:
\beq
\Phi=\frac{1}{\sqrt{2}}\left(\phi_1+ i\phi_2\right)\qquad 
\Phi^\dagger\Phi=\frac{1}{{2}}\left( \phi_1^2 + \phi_2^2\right)
\eeq
and the action becomes:
\bea
\label{twoU1}
\!\!\!\!\! S=\int \sqrt{-g} \left(g^{\mu \upsilon }\partial _{\mu }\Phi^\dagger
\partial _{\nu }\Phi-\frac{\alpha}{6}\Phi^\dagger\Phi R-W
\right).
\eea
The $K_\mu$ current and kernel become:
\beq
K_\mu = \partial_\mu K\qquad K=(1-\alpha)\Phi^\dagger\Phi.
\eeq
The inertial symmetry breaking implies that
dynamically $K\rightarrow \bar{K}$, hence:
\beq
\Phi^\dagger\Phi\rightarrow f^2/(1-\alpha)\qquad
f^2=2\bar{K}.
\eeq
However, now
this also defines the spontaneously broken
phase of the $U(1)$ symmetry,
We can therefore  rewrite: 
\beq
\Phi=\frac{f}{\sqrt{2}}\exp(i\tilde{\pi}/f+\sigma/f)).
\eeq 
In this representation we now have:
\bea
\label{twotwoU1}
\!\!\!\!\! S&=&\int \sqrt{-g} \left(g^{\mu \upsilon }e^{2\sigma/f}
\left(\frac{1}{2}\partial _{\mu }\sigma
\partial _{\nu }\sigma + \frac{1}{2}
\partial _{\mu }\tilde{\pi}\partial _{\nu }\tilde{\pi}\right)\right.
\nonumber \\
&& \qquad\left.
-\frac{\alpha}{12}e^{2\sigma/f}f^2 R
-\frac{f^4}{4}e^{4\sigma/f}\widehat{W}(\tilde{\pi}/f) \right).
\eea
where $\widehat{W}=e^{-4\sigma/f}W$.
The Weyl transformation is now implemented, as expected, by shifting $\sigma$
and rescaling the metric:
\beq
\label{form1}
\sigma\rightarrow \sigma + \epsilon f \qquad g_{\mu\nu}\rightarrow e^{-2\epsilon}g_{\mu\nu}
\eeq
while $\tilde{\pi}(x)$ and $f$ are held fixed. This is the usual view 
of the dilaton as a Nambu-Goldstone boson and the transformation
implemented on field variables. 

However, now we can see from
eq.(\ref{twotwoU1}) that an equivalent covariant global scale transformation is
also a symmetry:
\bea
\label{cov}
\sigma\rightarrow e^\epsilon\sigma\qquad \tilde{\pi}(x)\rightarrow e^\epsilon\tilde{\pi}(x)
\nonumber \\
f\rightarrow e^\epsilon f\qquad g_{\mu\nu}\rightarrow e^{-2\epsilon}g_{\mu\nu}.
\eea
Here we are doing a scale transformation on {\em all} dimensional quantities,
including the ``constant'' $f^2$ together with the metric transformation.

It is important to distinguish in a theory between
transforming dynamical quantities, \ie, fields, vs. transforming fixed input
parameters.  The latter is not a symmetry: only dynamical fields can
represent the symmetry group and have Noether currents.  
However, while $f$ appears to be a parameter of
the low energy (Einstein frame) theory, it is in fact 
a dynamical quantity, $f^2=2\bar{K}\propto \Phi^\dagger\Phi$
in the underlying (Jordan frame) theory, and eq.(\ref{cov}) represents the covariant 
Weyl transformation in that underlying theory.

Now, performing the metric redefinition, $g_{\mu\nu}\rightarrow e^{-2\sigma}g_{\mu\nu}$ 
as in the simple
single field example the action of  eq.(\ref{twotwoU1}) becomes:
\bea
\label{threeU1}
S&=&\int \sqrt{-g}g^{\mu \nu}\left( \frac{1}{2}\partial _{\mu }\sigma
\partial _{\nu }\sigma + \frac{1}{2}
\partial _{\mu }\tilde{\pi}\partial _{\nu }\tilde{\pi}\right)
\nonumber \\
&&\!\!\!\!\! \!\!\!\!\!
-\int d^4x\sqrt{-g}\left(\frac{\alpha}{12}f^2R
+\frac{f^4}{4}\widehat{W}(\tilde{\pi}(x)/f) \right).
\eea
The original Weyl invariance is hidden in the action of
eq.(\ref{threeU1}) since the new metric $e^{-2\sigma}g_{\mu\nu}$
has been made Weyl invariant under eq.(\ref{form1}).  However, the transformation
of eq.(\ref{cov})
remains nontrivial and reflects the Weyl
invariance of eq.(\ref{threeU1}).
Geometrically, the Weyl transformation rescales the radius of
the circle, $f$, while holding the angle variable $\tilde{\pi}(x)/f$ fixed.
While the transformation can be implemented by shifting $\sigma$, as in
eq.(\ref{form1}), it is 
convenient to treat $\sigma$ as transforming covariantly 
as in eq.(\ref{cov})
with the accompanying rescaling of $f$.

The $U(1)$ symmetry is also spontaneously broken.
The field $\tilde{\pi}/f$ has disappeared from the nonminimal gravitational 
coupling. It is the Nambu-Goldstone boson (NGB) of the broken $U(1)$ symmetry.
However, by contrast to the usual  Higgs mechanism, \eg, with a Mexican hat potential,
the Higgs boson is generally heavy with a mass given by $m^2\sim f^2 \lambda'$
where $\lambda'$ is the quartic potential defining the hat. 
Here the the Higgs boson is the dilaton, {\em but it is
massless as it too is a Nambu-Goldstone boson of broken
Weyl symmetry.} 

The model can be generalized by including more scalars.
With $N$ scalars and an $SO(N)$ invariant nonminimal coupling to gravity,
the inertial symmetry 
spontaneously breaks $SO(N)$ to $SO(N-1)$
by giving the kernel $\propto \sum\phi_i^2$ a VEV.
The fields are then confined to $S_{N-1}$ in theory space.
The inertial symmetry breaking
is independent of the potential, and the potential
can explicitly break this symmetry.
An arbitrary
point on $S_{N-1}$ will  dynamically move during the
inflationary phase. This motion is controlled by the potential,
$W(\phi_i)$.  It is interesting to speculate
that there maybe nontrivial topological configurations
associated with inertial symmetry breaking, in multiscalar theories,
corresponding to homotopy groups, such as $\Pi_3(SU(2))$, etc.

 Thus far we have considered only the inertial symmetry breaking.
Let us now consider the role of potential.
Here the
existence of flat directions is fundamentally important
if we want the final vacuum to have a small cosmological constant.
Perhaps the simplest potential in our original $(\phi_1,\phi_2)$ two
scalar model with a flat direction consists of a single field quartic interaction
\beq
W(\phi_1,\phi_2)= \frac{\lambda}{4}\phi_2^4=\lambda (\Phi-\Phi^\dagger{})^4
\eeq
where the flat direction is the $\phi_1$ axis.
An arbitrary field configutation after an initial
period of expansion is a
point on $S_1$. This will then dynamically evolve during the
inflationary phase toward the minimum. In the above example, an arbitrary point
on the $(\phi_1,\phi_2)$ circle will move toward the 
$\phi_2=0$ minimum of $W$.

In our $U(1)$ model, after performing the Weyl redefinition
to remove $e^{\sigma/f}$ factors, the potential is:
\beq
\label{flat}
\widehat{W} =\frac{\lambda f^4}{4}\sin^4(\tilde{\pi}/f).
\eeq
The flat direction corresponds to a shift symmetry,
$\Phi\rightarrow \Phi+\epsilon$ for real $\epsilon$.
The potential of eq.(\ref{flat}) breaks $U(1)\rightarrow Z_2$, since the potential
is invariant under shifts of 
$\tilde{\pi}(x)/f \rightarrow \tilde{\pi}(x)/f + \pi N $.
Quantum loops will necessarily lead to a running of the 
coupling $\lambda$ and modifications of the potential.

We can focus on the real scalar field theory Lagrangian in flat space:
\begin{equation}
L=\frac{1}{2}(\partial \tilde{\pi} )^{2}-\frac{1}{4}\lambda f^{4}\sin ^{4}\left(
\tilde{\pi} /f\right).  \label{S0}
\end{equation}%
To compute the potential, we introduce a classical source term in the 
lagrangian, $-J\tilde{\pi} $.  Through equations of motion,
$J$ induces the shift in the field,
\begin{equation}
\tilde{\pi} =\tilde{\pi} _{c}+\hbar ^{1/2}\hat{{\pi}}
\end{equation}%
where $\tilde{\pi} _{c}$ satisfies the renormalized equation of motion, $\partial
^{2}\tilde{\pi} _{c}+\lambda f^{3}\sin ^{3}\left( \tilde{\pi} _{c}/f\right) \cos \left(
\tilde{\pi} _{c}/f\right) +J=0$. The Lagrangian becomes:
\begin{equation}
L=L_{0}(\tilde{\pi} _{c})+[\hbar ]\hat{L}(\tilde{\pi} _{c},\widehat{\tilde{\pi} })
\end{equation}%
where, the on-shell $\tilde{\pi} _{c}$ with the source term 
leads to cancellation of tadpoles (linear $\hat{{\pi}}$ terms). Hence to $O(\hbar )$:
\begin{equation}
L_{0}(\tilde{\pi} _{c})=\frac{1}{2}(\partial \tilde{\pi} _{c})^{2}-\frac{1}{4}\lambda
_{r}f^{4}\sin ^{4}\left( \tilde{\pi} _{c}/f\right)   \label{Sc}
\end{equation}%
and: 
\begin{eqnarray}
\hat{L}(\tilde{\pi} _{c},\widehat{\pi }) &=&\frac{1}{2}(\partial \hat{\pi})^{2}-%
\frac{1}{2}G\hat{\pi}^{2}+\text{ }... 
\end{eqnarray}%
where:
\bea
G& =& \lambda f^{2}(3\sin ^{2}\left( \tilde{\pi} _{c}/f\right) -\sin
^{4}\left( \tilde{\pi} _{c}/f\right) )\nonumber \\
& & 
\sim
3\lambda f^{2}\sin ^{2}\left( \tilde{\pi} _{c}/f\right)
\eea
where we drop the subleading term for small $\tilde{\pi} _{c}/f$.
The effective theory is non-renormalizeable, but, like chiral
perturbation theory, it
becomes renormalizeable in an expansion
in large $f$, and we have displayed
the leading behavior for small fixed $\tilde{\pi} _{c}$ and large $f$.
This
expression, better than a polynomial expansion in $\tilde{\pi} _{c}/f,$ 
protects the global $Z$ shift symmetry $%
\tilde{\pi} _{c}/f\rightarrow \tilde{\pi} _{c}/f+2\pi N$. There is no
wave-function renormalization constant at one loop order. 

We integrate out the quantum fluctuations, $\hat{{\pi}}$. The effective
Lagrangian takes the form:\footnote{The effective action is: $-i\hbar \ln (Z)\sim -i\hbar \int \ln (1/(\ell
^{2}-\mu ^{2}))=i\hbar \int \ln (\ell ^{2}-\mu ^{2})$. The potential is 
$-i\hbar \int \ln (\ell ^{2}-\mu ^{2})$ .\ \ The
integral can be done by performing a Wick rotation ($\ell _{0}\rightarrow
i\ell _{0}$, $\ell ^{2}\rightarrow -\ell _{0}^{2}-\vec{\ell}^{2}$, and $%
d^{4}\ell \rightarrow id\ell _{0}d^{3}\ell $ ) and we use a Euclidean
momentum space cut-off, $\Lambda $.}
\bea
L &=&L_{0}(\tilde{\pi} _{c})-V_{eff}  \nonumber  \\
V_{reg} &=&i\hbar \int \frac{d^{4}\ell }{(2{\pi} )^{4}}\ln (\ell ^{2}-
3\lambda f^{2}\sin ^{2}\left( \tilde{\pi} _{c}/f\right)).
 \eea
  Let  
$y=\sin \left( \tilde{\pi} _{c}/f\right) $ 
and,
up to additive constants $\propto \Lambda ^{4}$ 
we obtain a regularized expression: 
\bea
V_{reg}&=& \frac{1}{32\pi ^{2}}\left(
(3\lambda f^{2}y^2)\Lambda ^{2}
\right.
\nonumber \\
&&\!\!\!\!\!  \!\!\!\!\!  \!\!\!\!\! \!\!\!\!\! \left. -\frac{1}{2}
(3\lambda f^{2}y^{2})^2\left( \ln \left( \frac{\Lambda ^{2}}{3\lambda f^{2}y^2}%
\right) +\frac{1}{2}\right) \right) .
\eea
Not surprisingly, we have quadratic and log divergences. 
It doesn't
matter how we define $\Lambda$ under Weyl transformations,
as this is only a regularized expression.

We now introduce a set of three counterterms:
\bea
V_{ct} &=&\delta _{0}f^{4}+\frac{1}{2}\delta _{1}f^{4}\sin ^{2}\left( \tilde{\pi}
_{c}/f\right)\nonumber \\
& & \qquad +\frac{1}{4}\delta _{2}f^{4}\sin ^{4}\left( \tilde{\pi} _{c}/f\right).
\eea
The counterterms are defined by imposing renormalization conditions.
With a
vanishing $\sin^{2}(\tilde{\pi}/f)$ term we will 
require a definition of the quartic countertem
at a nonzero scale $y=\sin(\tilde{\pi}_c/f)=c$, due to the infrared singularity.
Note that this is a Weyl invariant specification since
$\tilde{\pi}_c/f$ is Weyl invariant, by eq.(\ref{cov}).
The counterterms are determined by:
\bea
0&=& \left( V_{reg}+f^{4}\delta _{0}\right) _{y=0} 
\nonumber \\
0&=& \left( \frac{\partial ^{2}}{\partial y^{2}}V_{reg}+f^{4}\delta _{1}\right)_{y=0}
\nonumber \\
0&=&\left(\frac{\partial ^{4}}{\partial y^{4}}V_{reg}+3!f^{4}\delta _{2}\right)_{y =c}.
\eea
From the regularized expression we have:
\bea 
\label{A}  \delta _{0}& =& 0
\nonumber  \\
\delta _{1} & = & -
\frac{1}{16\pi ^{2}f^{2}}\left( 3\lambda \Lambda ^{2}\right)
\nonumber \\
\delta _{2}&=& -\frac{9}{16\pi ^{2}}
\left( \lambda ^{2}\ln \frac{\Lambda ^{2}}{
3\lambda f^{2}c^{2}}-\frac{\allowbreak 11}{3}\lambda ^{2}\right).
\nonumber
\eea
Hence the renormalized potential is:
\bea
\label{final}
V_{reg}+V_{ct}&=&-\frac{1}{4}\lambda
_{r}f^{4}\sin ^{4}\left( \tilde{\pi} _{c}/f\right)\nonumber\\
&&   \!\!\!\!\!   \!\!\!\!\!   \!\!\!\!\!   \!\!\!\!\!  \!\!\!\!\!
\!\!\!\!\!   \!\!\!\!\!   \!\!\!\!\!  \!\!\!\!\!   
-\frac{9f^{4}}{64\pi ^{2}}\left[ 
\lambda^2 _{r}\sin ^{4}\left( \tilde{\pi} _{c}/f\right) \left( \ln
\left( \frac{\sin ^{2}\left( \tilde{\pi} _{c}/f\right) }{c^2}\right) -\frac{25}{6}%
\right) \right]. \nonumber \\
&& 
\eea
This 
reproduces the usual CW result for $\lambda \phi^4$ theory
with $\tilde{\pi}_{c}\rightarrow f\sin(\tilde{\pi} _{c}/f)$.

The leading log limit is of the 
form:
\bea
\label{final2}
V
&=&   
-\frac{9f^{4}\lambda^2}{64\pi ^{2}} 
\sin ^{4}\left( \tilde{\pi} _{c}/f\right) \ln
\left( \frac{\sin ^{2}\left( \tilde{\pi} _{c}/f\right) }{c^2}\right)%.
\eea
and the sub-leading (non-log) terms can be subsumed into the definition
of $c^2$.
This provides a calculation
of $\beta_1$ (see \cite{CTH} for discussion
of the direct RG approach to Coleman-Weinberg potentials).

The results of eqs.(\ref{final},\ref{final2}) are Weyl covariant 
quantum loop induced potentials, and lead
to invariant actions, $\sim\int \sqrt{-g}\; V$, by the transformation of 
eq.(\ref{cov}). There is no requirement of Weyl invariance in
the intermediate steps of the regularized calculation.

\section{Conclusions}

We have discussed  how inflation and
Planck scale generation emerge as a unified phenomenon from a dynamics associated with
global Weyl symmetry  \cite{shapo}\cite{GB}\cite{authors}\cite{FHR}\cite{Ferreira:2018itt}
\cite{FHR2}\cite{FHRN}. 
We have placed particular emphasis 
upon the Weyl current, $K_\mu$.  In the pre-inflationary universe,
the Weyl current density, $K_0$, is driven to zero by general expansion.  
However,   $K_\mu$  has a kernel structure, \ie, $K_\mu =\partial_\mu K$,
and as   $K_0\rightarrow 0$,  the kernel evolves as  $K\rightarrow \bar{K}$ constant.  
This resulting constant $\bar{K}$ is the order parameter
of the Weyl symmetry breaking, indeed, $\bar{K}$ directly defines $M_P^2$.

This defines ``inertial spontaneous symmetry breaking'' \cite{Ferreira:2018itt}.
This mechanism involves a new form of dynamical scale symmetry breaking.
driven by the formation of a nonzero kernel, $\bar{K}$. The scale breaking 
has nothing to do with any
potential in the theory, but is soley dynamically generated by gravity.
In addition,
a scale invariant  potential with a flat-direction ultimately determines 
the  vacuum of the theory and relative VEV's of the scalar fields contributing to $\bar{K}$.
There is a harmless dilaton associated with the dynamical symmetry breaking
which decouples from everything except gravity \cite{FHR2}.  

We  illustrated this phenomenon in  a single scalar field theory, $\phi $,
with nonminimal coupling. We see how, starting in the Jordan frame, we smoothly evolve
into the Einstein frame. 
The theory has a conserved current, $K_\mu=(1-\alpha)\phi \partial_\mu \phi$.
The scale current charge density
dilutes to zero in the pre-infationary phase
$K_{0}\sim (a(t))^{-3}$. The kernel, $K=(1-\alpha)\phi^2/2 $, and hence
the VEV of $\phi$, are driven to 
constants. With $\alpha<0$, this 
induces a positive Planck (mass)$^2$. 
The resulting inflation in this simple model is eternal.

In multi-scalar-field theories we see that we will have a
generalized $K=\sum_i(1-\alpha_i)\phi_i^2/2$. As this is driven to a constant, it
defines an ellipsoidal constraint on the scalar field VEV's, and
the Planck scale is again generated $\propto \bar{K}$.
An inflationary slow-roll is then associated with the field VEV's migrating along the ellipse,
ultimately flowing to an intersection of the flat direction
with the ellipsoid. In a  two-scalar scheme
the terminal phase of inflation is similar to standard $\phi^4$ inflation, since the effective
theory is now essentially Einstein gravity with a fixed $M^2_P$.
The final cosmological constant vanishes
with exact flatness of the potential.

Any Weyl symmetry breaking effect is intolerable in these schemes and will
show up as a nonzero divergence in the $K_\mu$ current (a Weyl anomaly). 
We show how a Weyl invariant Coleman-Weinberg action is computable
when renormalization masses are defined by field (VEV's) contained in the action.
It is natural to renormalize the theory with the scale $f^2=2\bar{K}$ itself.

We demonstrated in this calculation how a $U(1)$ symmetry is broken solely by 
the inertial symmery breaking mechanism.  The $U(1)$ scalar ends up
in a Higgs phase, but the Higgs boson is now the massless dilaton.
There is considerable work to continue along these lines exploring how
gauge symmetries can be broken, how topological vacua might arise, and 
possible applications to the real world. 
 
There remain challenges. For example, if one imbeds this
in ``A-gravity'' \cite{Strumia} then the 
$\alpha$ terms can
feedback, via quantum loops, upon gravity
on scales above $M_{P}$, and, \eg,
violate a flat direction in perturbation theory.
This poses
a potential problem for the program as it leads to an
uncontrollably large 
induced $U(1)$ invariant potential $\sim \lambda' (\Phi^\dagger \Phi)^2$,
which, in turn, would lead to an uncontrollably large cosmological constant in the
Einstein frame theory.  

Our philosphy here, however, may differ from that of A-gravity.
We view the $R^2$ terms to be induced, in analogy to Coleman-Weinberg potentials, 
and therefore, taking a
form schematically like $\sim \epsilon R^2\log(K/R)$ etc. As such, we wouldn't
treat these terms as part of the propagator, \ie, we would not include 
$1/\epsilon p^4$ terms for small, perturbative $\epsilon$.
Our theory is then effectively cut-off at the Planck scale because the
$M_P^2 R$ term becomes an interaction $\alpha \phi^2 R$
at trans-Planckian energies. Gravity might be viewed as a collective
phenomenon in this view.  Alternatively, we may seek resolution with
A-gravity by inclusion of \eg, the Weyl photon, or other
generalizations of the present models.

Possible resolutions to these and other issues
are currently under study and will be discussed elsewhere.

%\vspace{1.0in}

\vskip 0.5 in
\noindent
 {\bf Acknowledgements}

\vspace{0.1in}
We thank W. Bardeen, Pedro Ferreira, Johannes Noller, and Graham Ross 
for discussions.
This work was done at 
Fermilab, operated by Fermi Research Alliance, 
LLC under Contract No. DE-AC02-07CH11359 with the United States 
Department of Energy.

\end{document}